\documentstyle[12pt]{article}
\newcommand{\tr}{\mbox{tr}}

\begin{document}
\baselineskip=15.5pt

\begin{titlepage}

\begin{flushright}
hep-th/0009017\\
PUPT-1950\\
\end{flushright}
\vfil

\begin{center}
{\huge Stable Massive States\\}
\vspace{3mm}
{\huge in 1+1 Dimensional NCOS}
\end{center}

\vfil
\begin{center}
{\large Christopher P. Herzog and Igor R. Klebanov}\\
\vspace{1mm}
Joseph Henry Laboratories, Princeton University,\\
Princeton, New Jersey 08544, USA\\
\vspace{3mm}
\end{center}

\vfil

\begin{center}
{\large Abstract}
\end{center}

\noindent
We study the case of open string theory on a D1-brane in a critical 
electric field.
It was argued in hep-th/0006085 that the massless open string modes of this 
theory decouple from the massive modes, corresponding to a decoupling of the 
U(1) degrees of freedom in the U(N) gauge theory dual.  To 
provide further support for the decoupling, we present several examples of open 
string disk amplitudes.  Because of the 
decoupling, many of the massive open string states are stable, indicating the 
presence of a number of correspondingly stable states in the gauge theory.  We 
provide a lengthy list of these stable states.  However, when the theory is 
compactified in the spatial direction of the electric field, 
we demonstrate that sufficiently massive such states may decay into wound 
closed strings. 
\vfil
\begin{flushleft}
September 2000
\end{flushleft}
\vfil
\end{titlepage}
\newpage


\section{Introduction}

Non-commutative open string theory (NCOS) \cite{Seiberg, Gopak}
has been a subject of growing interest 
lately. This is 
the theory of open strings living on a D-brane in 
the presence of a critical electric field, which causes an infinite 
rescaling of the string tension and coupling.  NCOS is in 
many ways very similar to the usual 
open string theory.  Two important differences are, one, the amplitudes 
contain additional phase factors distinguishing different
orderings of the vertex operators, and, two, 
the open string sector completely decouples from closed strings with zero 
winding number.

NCOS can live on a brane of arbitrary dimension, but we shall focus in this note 
on the case of the D-string.  The electric field on a D-string is quantized:  
N units of electric flux describe a bound state of one 
D-string with N fundamental strings \cite{Witten}.  
S-duality interchanges D-strings with 
fundamental strings, and hence, by S-duality, NCOS on a D-string is 
related to a U(N) Super Yang Mills (SYM) 
theory in a sector with one unit of electric 
flux~\cite{Gukov, Verlinde, Gopak2,Klebanov}.  
Furthermore, the gauge group U(N) factors into a 
product U(1)$\times$SU(N)/$Z_N$.  The U(1) factor is free, and in this way 
duality predicts that there must be corresponding decoupling in the NCOS~\cite{Klebanov}.  

The authors of \cite{Klebanov} showed that it is the massless open strings which 
decouple, and they presented a general argument involving the structure of the 
Green's functions.  Here we show that, in the absence of a tachyon, 
a consequence of the decoupling of the massless modes is that a 
large class of the massive modes become stable.  All decay channels involving 
massless scalars are forbidden, and for many massive modes, 
these decay channels 
are the only kinematically accessible ones. 

However, when the theory is compactified along the spatial direction of the 
D-string, it is possible to create wound closed strings.  These wound strings, 
unlike the strings with zero 
winding number, do not decouple from the open string sector.  In fact, 
the wound strings provide an additional decay channel for the massive 
open strings.  We show that the leading SO(8) Regge trajectory states
(the states that carry maximum SO(8) spin for a given mass level),  
which would otherwise be stable, can decay into wound closed strings 
provided that energy conservation can be satisfied.  As the mass 
of the lightest wound strings is proportional to the radius of 
compactification, we suspect that all of the stable states 
we find can decay into wound closed strings if the compactification 
radius is made small enough.

In support of their general Green's function argument for decoupling of the 
massless open strings, the 
authors of \cite{Klebanov} present three disk amplitude calculations: 
the four-point amplitude for massless NS open 
superstrings, the forward scattering of a massless scalar off a tachyon in 
bosonic string theory, and the bosonic 4-tachyon amplitude.  In this note, we 
tie up a loose end by considering the backward scattering of a tachyon off a 
massless scalar in bosonic string theory.  We also present two superstring 
calculations involving a massive NS state.  In particular we consider the $l=1$ leading Regge trajectory state.  We 
calculate the two massive-two NS massless scalar amplitude and also the 
four-point massive amplitude.  Because these open string amplitudes require less formalism and are easier to understand, we will present them before discussing amplitudes involving wound strings.  But first, we begin with a discussion of the stable, massive open string modes.

\section{Stable Massive Modes}

In any scattering event, energy conservation must hold.  Moreover, the 
mass level of an open superstring is given by $\alpha'_e m^2 = l$ where $l$ is a 
non-negative integer.  For a general n-body 
decay, energy conservation means 
\begin{equation}
\sum_{i=1}^n \sqrt{l_i} \leq \sqrt{l}
\label{econs}
\end{equation}
where the letter 
$i$ indexes the product states.  In classifying the stable massive modes, we begin with an almost 
trivial observation.  As the massless modes have decoupled, the only allowed 
decay products have an $l$ of at least one, and the lowest energy decay state 
will consist of two such open string modes.  Thus, by conservation of energy, 
all states with $l<4$ are stable.  When an $l=4$ state decays, the two 
$l=1$ modes are produced at threshold.  In more than two dimensions, the absence of phase space for the final state would mean the $l=4$ states are stable.  However, in two dimensions, we can only say that, as the decay products have no energy to separate, we can never tell whether the $l=4$ state has or has not decayed.  With this caveat in mind, we will say that the $l=4$ states are stable as well.

To proceed further in our classification of stable massive modes, we 
consider angular momentum.  A more limited form of our argument 
was employed in \cite{Bala}.  We need some nontrivial results from 
representation theory, and for background, we refer the reader to 
\cite{FultonHarris}.  
In 9+1 dimensions, massive open string modes 
fall into irreps of SO(9).  The classical irreps of SO(9) 
and also of SO(8) are described by Young tableaux 
with four rows.  In what follows, we will ignore spinor representations.  
Following the notation of \cite{FultonHarris}, we label Young tableaux with the 
Greek letters $\mu,\nu,\ldots$ and irreps by $\Gamma_\mu, \Gamma_\nu, \ldots$.  
The $j$th row of a Young tableau $\nu$ has $\nu_j$ boxes.

Consider an n-body decay of the state with Young tableau $\mu$ into states with 
Young tableaux $\nu \in P$ where all of the Young tableaux 
correspond to irreps of SO(9).  
In general, the relative motion of the decay products will transform under a 
representation of SO(9) which we will call $R$.  
Angular momentum conservation becomes the 
necessity that the irrep $\Gamma_\mu$ be included in the tensor product of the 
$\Gamma_{\nu}$ and of $R$:
\[
\Gamma_\mu \in \left( \bigotimes_{\nu \in P} \Gamma_{\nu} \right) \otimes R.
\]

However, $R$ can be ignored.  
Our open strings are confined to a D-string.  As a result, the relative motion 
of the decay products will not produce any angular momentum in the directions 
transverse to the D-string.  This fact suggests that, given an open string mode 
with angular momentum in an irrep of SO(9), we should decompose it into irreps 
of SO(8) transverse to the D-string.  $R$ is trivial under this SO(8), 
and we can use the angular momentum condition 
\begin{equation}
\Gamma_\mu \in \bigotimes_{\nu \in P} \Gamma_{\nu}.
\label{amcons}
\end{equation}
where now $\Gamma_\mu$ and $\Gamma_\nu$ correspond to irreps of SO(8).

Let us demonstrate how this decomposition from SO(9) into SO(8) 
works in a simple case.  The first excited NS sector state, ignoring ghost insertions, has a 
vertex operator in the minus one picture,   
\[
V = \zeta_{\mu\nu} \psi^\mu \partial X^\nu e^{ik\cdot X}
\]
where $\zeta$ is in a traceless, symmetric representation of SO(9).  
Specifically, this state corresponds to a Young tableau with only one row and 
with two boxes and is an example of a state in the leading Regge trajectory.  As representations of 
SO(8), the vertex operator decomposes into\footnote{
Our indexing conventions are that $\alpha,\beta,\ldots \in (0,1)$ and 
$i,j,\ldots \in (2,3, \ldots ,9)$.}
\begin{eqnarray*}
V_{t} &=& \zeta_{ij} \psi^i \partial X^j e^{ik\cdot X} \\
V_{v} &=&  {\alpha'_e}^{1/2} \bar{k}_\alpha \xi_i 
(\psi^\alpha \partial X^i + \psi^i \partial X^\alpha) e^{ik\cdot X} \\
V_{s} &=& ({\alpha'_e} \bar{k} \cdot \psi \bar{k} \cdot \partial X -
\frac{1}{8} \delta_{ij} \psi^{i} \partial X^{j}) e^{ik \cdot X}
\end{eqnarray*}
where $\bar{k}^\alpha = \epsilon^{\alpha \beta} k_\beta$.
The first NS massive state has decomposed into SO(8) irreps corresponding to 
Young tableaux with only one row and with two, one, and zero boxes.  Indeed, 
this pattern will continue for the higher 
leading Regge trajectory states.  In general, a 
leading Regge trajectory state at mass level $l$ corresponds to a Young tableau 
$\nu$ such that $\nu_1=l+1$ and $\nu_j=0$ for $j>1$.  The 
leading Regge trajectory state at mass level $l$ will decompose under SO(8) into 
Young tableaux with one row and with $l+1, l, \ldots ,0$ boxes.  Each such 
tableau will appear only once in the decomposition, as can be verified by 
checking the dimensions.

For more general irreps of SO(9), there is a similar known decomposition.  Let 
$\Gamma_\lambda$ be an irrep of SO(9).  Under SO(8), $\Gamma_\lambda$ decomposes 
into $\bigoplus \Gamma_{\bar\lambda}$ where
\[
\lambda_1 \geq \bar\lambda_1 \geq \lambda_2 \geq \bar\lambda_2 \geq 
\lambda_3 \geq \bar\lambda_{3} \geq \lambda_4 \geq |\bar\lambda_4|. 
\]
The $\lambda_i$ and $\bar\lambda_i$ are all integer or, for spinor 
representations, all half integer.

Therefore, to summarize what we should do to 
test if a particular decay is allowed by angular 
momentum conservation is to find out which irreps of SO(9) the original state 
and the decay products transform under.  Next, we decompose these irreps into 
irreps of SO(8) according to the 
prescription given above.  We note that the 
relative motion of the decay products cannot produce any SO(8) angular momentum, 
and so, we invoke condition (\ref{amcons}).  
However, in what follows, we will not try to squeeze 
every last drop of information from this condition.  Rather we will only 
consider what happens to the first row of 
the SO(8) Young tableaux under the tensor product.  
As a result, we will obtain a lengthy although probably incomplete 
list of stable states.  For convenience, we relabel the length of 
the first row $\lambda_1 \equiv J$.  
The condition on the first row that we will prove now 
and use later is that 
\begin{equation}
J \leq \sum_{i=1}^n J_i,
\label{amcons2}
\end{equation}
where the $J$ without a subscript corresponds to the original open string state, and the $i$ index the decay products.  

In general, we expect to be 
able to decompose the tensor product of two irreps of any simple Lie group (and 
indeed of many other types of group as well) into a sum of other irreps
\[
\Gamma_\lambda \otimes \Gamma_\mu = \bigoplus_\nu N_{\lambda\mu\nu} \Gamma_\nu.
\]
R.~C.~King \cite{RCKing} has shown that, ignoring spinor representations, for 
both the symplectic and orthogonal groups, the multiplicities 
$N_{\lambda\mu\nu}$ are given by the formula
\[
N_{\lambda\mu\nu} = \sum_{\zeta,\sigma,\tau} M_{\zeta\sigma\lambda}
M_{\zeta\tau\mu} M_{\sigma\tau\nu}
\]
where $M_{\lambda\mu\nu}$ is the corresponding multiplicity for the general 
linear group.  A standard way of calculating $M$ is to use the 
Littlewood-Richardson rules.  Suppose that the Young tableau 
$\nu$ has $J_\nu$ boxes in the first row.  It follows from these
rules that for $M_{\lambda\mu\nu}$ to be nonzero the condition 
$\mbox{max}(J_\lambda, J_\mu) \leq J_\nu \leq J_\lambda + J_\mu$ must hold. 
Then in our formula for $N$, for a term in the sum to 
be nonzero, $J_\sigma \leq J_\lambda$ and $J_\tau \leq J_\mu$, and also,
$J_\sigma + J_\tau \geq J_\nu$.  Putting these three inequalities together, 
we find that for $N_{\lambda\mu\nu}$ to be nonzero,
$J_\nu \leq J_\lambda + J_\mu$.  
We have proven the desired result for the case of 2-body decay.  
The extension to n-body decay should be clear and follows by induction. 

Condition (\ref{amcons2}) is most easily satisfied when the right hand
side is as large as possible, i.e. when the first rows of the Young tableaux of the decay products are as long as possible.  The leading SO(8) Regge trajectory states, by which we mean the first term in the SO(8) decomposition of the leading SO(9) Regge trajectory states, maximize this length at a given mass level $l_i$.  Therefore, we set $J_i = l_i + 1$ and in its final 
form, angular momentum conservation for us will mean that  
\begin{equation}
\sum_{i=1}^n l_i \geq J-n.
\label{alcons}
\end{equation}
In what follows, we use $s=l+1-J$ instead of $J$.

First, let us consider the case $s=0$.  Squaring the conservation of energy 
inequality (\ref{econs}) and adding the condition from conservation of angular momentum (\ref{alcons}), we 
find
\[
\sum_{i<j} \sqrt{l_i l_j} \leq \frac{n-1}{2}.
\]
As massless decay products 
are not allowed, $l_i \geq 1$.  Thus, the left hand side of the above inequality is at least $n(n-1)/2$, 
and moreover $n\geq 2$.  The inequality cannot be satisfied, indicating that all states with $s=0$ are stable.  
In other words, the leading SO(8) Regge trajectory state is always stable.

Now let us consider the decay of a state $(l>4,s)$ 
into $n$ other open string modes.  
Note first that if our two inequalities (\ref{econs}) and (\ref{alcons}) 
allow a $n>2$ body decay, 
then an $n-1$ body decay is allowed as well.  
In particular, let two of the decay products be $l_1$ and $l_2$.  We can 
always replace these two states with a third state $l_3$ such that 
$\sqrt{l_1} + \sqrt{l_2} \geq \sqrt{l_3}$ but also such that 
$(l_1+1) + (l_2 + 1)\leq l_3+1$.  Therefore, we can restrict 
our analysis to two body decays.

For two body decay, the two inequalities are $\sqrt{l_1} + \sqrt{l_2} \leq 
\sqrt{l}$ and $s \geq l - l_1 - l_2 -1$.  To get a lower bound for $s$, we must 
maximize $l_1 + l_2$.  We choose $l_1 \leq (\sqrt{l} - 1)^2$ and $l_2 = 1$, and 
find that, when $l > 4$, all states with 
\begin{equation}
s < 2 \sqrt{l} - 3
\label{scond}
\end{equation} 
are stable.  In the case of equality, the state $(l, s=2 \sqrt{l} -3)$ is 
stable because the decay products cannot separate.  One interesting consequence is that for $s=1$, all states are stable.  

Here are two kinds of SO(9) irreps which, when decomposed, produce $s=1$ states.  The first irrep is a leading Regge trajectory state.  When we decompose the leading Regge trajectory into SO(8), the second term in the sum, the term with $J=l$, has $s=1$.  The other kind of SO(9) irrep is the first subleading Regge trajectory, which is described by the Young tableau $(\nu_1 = l, \nu_2=1, \nu_3=1, \nu_4=0)$.

We can go on in this fashion.  For example, for $s=2$, 
only the states with $l=5$ and 6 are possibly unstable.  Note that in the case 
of the leading Regge trajectory, condition (\ref{scond}) is somewhat 
stronger because under SO(8), the leading Regge trajectory states decompose into 
Young tableaux with only one row.

\section{Three Open String Disk Amplitudes}

Before describing specific amplitudes, we take some time to describe how the 
electric field modifies what otherwise would be a completely straighforward and 
routine calculation in string perturbation theory.  The modifications due to the 
electric field and the D-brane can be found in various places in the literature 
\cite{FT, Gukov, SeiWit}, but for reasons of clarity, 
we would like to present a brief summary here.

We turn on an electric field in the 01 directions with strength $E$.
In the standard NCOS prescription, the 
string tension in the directions transverse to the D1-brane is set to 
$T_e = 1/(2\pi \alpha'_e)$ while $T = 1 /(2\pi \alpha')$ is left as the tension 
in the 01 directions where $\alpha' = (1-E^2) \alpha'_e$.  
One can show, for example from considering the Born-Infeld action, that 
the open string coupling constant
is related to the usual closed string coupling $g$ through 
$G_o^2 = g \sqrt{1-E^2}$.  

In the presence of an electric field, one would think that the open 
string vertex operators become modified.  However, as long as we consider 
only open string scattering events, it is possible to work with the zero 
field vertex operators and absorb most of the effect of the electric field into 
the Green's functions.  The one modification that we need to make to the vertex 
operators is to replace every $\alpha'$ with $\alpha'_e$.
The boundary Green's functions are then:
\begin{eqnarray}
\langle X^\alpha(y) X^\beta(y') \rangle &=& 
-2 \alpha'_e \eta^{\alpha\beta} \ln |y-y'| + i \frac{\theta}{2} \epsilon^{\alpha \beta} \mbox{sgn}(y-y')
\label{XXGreens} \\
\langle \partial X^\mu(y) \partial X^\nu(y') 
\rangle &=& -\frac{\alpha'_e}{2} \frac{\eta^{\mu\nu}}{(y-y')^2} \\
\langle \psi^\mu(y) \psi^\nu(y') \rangle &=& \frac{\eta^{\mu\nu}}{y-y'}
\end{eqnarray}
where $\theta = 2 \pi \alpha'_e E$.
In the NCOS limit, $E \rightarrow 1$ and $\alpha' \rightarrow 0$ but 
$\alpha'_e$ and $G_o$ are held fixed and finite \cite{Seiberg,Gopak}.  
In the following sections we will confirm that it is the extra phases 
coming from (\ref{XXGreens}) which lead to the 
decoupling of the massless modes in the NCOS limit.

\subsection{Bosonic two tachyon-two massless scalar amplitude}

The tachyon and massless scalar vertex operators are respectively
\[
V_T = G_o e^{i k \cdot X} \;\;\;\;\; \mbox{and} 
\;\;\;\;\; V_S = \frac{G_o}{\sqrt{\alpha'_e}}
\zeta \cdot \partial X e^{i p \cdot X}.
\]
The mass shell conditions are $k^2 = 1/\alpha'_e$, $\, p^2 = 0$, and $p \cdot 
\zeta = 0$. Because we have restricted the ends of the open string to a 
D-string, $\zeta$ cannot have any component in the zero or one direction.  
This fact leads to simplification in the calculation 
of the amplitude --- we need only contract 
the $\exp(ik\cdot X)$ and $\exp(ip\cdot X)$ operators among themselves.

The vanishing of the forward scattering of a scalar off a tachyon was calculated 
in \cite{Klebanov}, so we will look at backward scattering and choose the center 
of mass frame momenta to be
\[
k_1 = \left( \begin{array}{c} e \\ p \end{array} \right) \;\;\;
k_2 = \left( \begin{array}{c} p \\ -p \end{array} \right) \;\;\;
k_3 = \left( \begin{array}{c} -e \\ p \end{array} \right) \;\;\;
k_4 = \left( \begin{array}{c} -p \\ -p \end{array} \right)
\]
where $e^2-p^2 = -1/\alpha'_e$.  We choose
\begin{eqnarray*}
s &=& -\alpha'_e(k_1 + k_2)^2 \\
t &=& -\alpha'_e(k_1 + k_3)^2 \\
u &=& -\alpha'_e(k_1 + k_4)^2.
\end{eqnarray*}
There are six possible orderings of the four vertex operators on the disk.  The 
orderings pair up into the $tu$, $su$, and $st$ channels.  For a given pair, the 
phases created by the Green's function (\ref{XXGreens}) are equal and opposite, 
resulting in a cosine function.  Specifically, the phases are
\begin{eqnarray*}
tu: & & \cos \left[ 
\pi \alpha'_e E (k_1 \wedge k_3 + k_2 \wedge k_4) \right] \\
su: & & \cos \left[ 
\pi \alpha'_e E (k_1 \wedge k_2 + k_3 \wedge k_4) \right] \\
st: & & \cos \left[ 
\pi \alpha'_e E (k_1 \wedge k_3 - k_2 \wedge k_4) \right].
\end{eqnarray*}
It is these phases which produce the new and interesting behavior in the NCOS 
limit.
Summing over the three channels, the total amplitude is given by
\begin{eqnarray}
\lefteqn{\frac{G_o^2}{\alpha'_e} \, \delta^2 \left( \sum_i p_i \right) 
\zeta_1 \cdot \zeta_2 \left( B(-u, -s) + \right.} & & \nonumber\\
& & \left. B(-1-t,-s) \cos[(s+1)\lambda] +
B(-1-t,-u) \cos[(u+1)\lambda] \right)
\end{eqnarray}
where $2 \alpha'_e \lambda \equiv \theta$.  The Beta functions are defined such 
that $B(x,y) = \Gamma(x) \Gamma(y) / \Gamma(x+y)$.
In the critical field limit $\lambda = \pi$, it is not too 
difficult to see that the amplitude vanishes.  The two crucial facts used in the 
calculation are $s+t+u = -2$ and the Gamma function identity 
\[
\Gamma(x) \Gamma(1-x) = \frac{\pi}{\sin(\pi x)}.
\]

\subsection{Massive Superstring Disk Amplitudes}

As mentioned in the Introduction, we will consider the 
$l=1$ leading Regge trajectory 
NS massive mode.  The zero and minus one picture vertex operators for 
this mode are
\begin{eqnarray}
V_{-1} &=& \frac{G_o}{\sqrt{\alpha'_e}}
e^{-\phi} \zeta_{ij} \psi^i \partial X^j e^{ik \cdot X} 
(y).
\nonumber \\
V_{0} &=& \frac{G_o}{\alpha'_e} \zeta_{ij} \left( 
\partial X^i \partial X^j - 
\alpha'_e \psi^i \partial \psi^j + i \alpha'_e (k \cdot \psi ) \psi^i \partial 
X^j \right) e^{ik \cdot X} (y).
\end{eqnarray}
For simplicity, we are only considering the vertex operator which 
is a traceless symmetric tensor under SO(8).  
The mass shell condition is $k^2 = -1/\alpha'_e$. 

We consider first the 4-massive amplitude.  The center of mass frame momenta are
\[
k_1 = \left( \begin{array}{c} e \\ p \end{array} \right) \;\;\;
k_2 = \left( \begin{array}{c} e \\ -p \end{array} \right) \;\;\;
k_3 = \left( \begin{array}{c} -e \\ -p \end{array} \right) \;\;\;
k_4 = \left( \begin{array}{c} -e \\ p \end{array} \right).
\]
There are a bewildering number 
of possible contractions to consider.  In performing the calculation, 
we chose to put particles three and four in the zero picture.  
Also, making use of the $SL_2(R)$ symmetry, we sent 
$y_3 \rightarrow \infty$.  Working through the combinatorics, 
the algebra, and the integrals, it turns out that only one of the 
contractions contributes to the amplitude.  Specifically, it is a contraction 
involving the third term in each of the zero 
picture vertex operators.  To cut a 
long story short, the amplitude simplifies to
\begin{equation}
{\mathcal A} \sim \frac{G_o^2}{\alpha'_e} \, \delta^2 \left(\sum_i k_i \right) \pi (s-2) 
\frac{\cos(\pi s) - \cos(s \lambda \frac{p}{e})}{\sin(\pi s)}
\tr(\zeta_1 \cdot \zeta_3) \tr(\zeta_2 \cdot \zeta_4). 
\end{equation}
The amplitude is similar to the bosonic 4-tachyon amplitude considered in 
\cite{Klebanov}.  As there are no massless particles involved, neither
amplitude vanishes at critical electric field, $\theta = 2 \pi \alpha'_e$.  
However, in this NCOS limit, both amplitudes 
do have a much softer UV behavior than the corresponding $\theta=0$
amplitudes.  Indeed, for large $s$, we may write:
\[
\cos (\pi s) - \cos \left( \pi s \frac{p}{e} \right) = 
\cos (\pi(s-2))- \cos \left( \pi \sqrt{s(s-4)} \right) \approx 
- 2\pi {\sin \left [\pi(s-2) -\pi (s-2)^{-1} \right ]\over s-2}
\ .
\]
Therefore, for large $s$,
\[
{\mathcal A} \sim -\frac{2 \pi^2 G_o^2}{\alpha'_e} 
\; \delta^2 \left(\sum_i k_i \right)
{\sin \left [\pi(s-2) -\pi (s-2)^{-1}\right ]\over \sin [\pi(s-2)]}
\tr(\zeta_1 \cdot \zeta_3) \tr(\zeta_2 \cdot \zeta_4) 
\]
which implies that the residues of the poles at $s=n$ fall 
off as $1/(n-2)$ while far from the poles, the amplitude is a constant.  
The softer behavior is not altogether surprising because we 
expect highly relativistic particles and massless particles to 
behave similarly. In \cite{Klebanov} it was argued that the softened
high-energy behavior is consistent with expectations based on 
perturbative gauge theory. 

To conclude this subsection, and to provide additional 
support for the decoupling of the massless modes, 
we consider the forward and backward scattering of 
the SO(8) tensor NS massive mode off of a massless scalar.  
Because of the general argument presented in \cite{Klebanov}, 
the corresponding SO(8) vector and singlet massive modes should
decouple from the massless scalar as well, but we will not check this fact.
The NS sector massless 
scalars have zero and minus one picture vertex operators 
\begin{eqnarray*}
V_{-1} &=& G_o e^{-\phi} \xi \cdot \psi e^{i k \cdot X} (y) \\
V_0    &=& \frac{G_o}{\sqrt{\alpha'_e}} \xi_\mu \left( \partial X^\mu + 
i \alpha'_e k \cdot \psi \psi^\mu \right) e^{i k \cdot X} (y).
\end{eqnarray*}
As for the bosonic massless scalar, 
$\xi \cdot k = 0$, $k^2 = 0$, and, as the open string is restricted to a 
D-string, $\xi$ can have no components in the $01$ directions.  
We chose to do our calculations with both massive 
vertex operators in the minus one picture.

We consider first the case of forward scattering:
\[
k_1 = \left( \begin{array}{c} p \\ p \end{array} \right) \;\;\;
k_2 = \left( \begin{array}{c} e \\ -p \end{array} \right) \;\;\;
k_3 = \left( \begin{array}{c} -p \\ -p \end{array} \right) \;\;\;
k_4 = \left( \begin{array}{c} -e \\ p \end{array} \right).
\]
Because in forward scattering $t=0$, only contractions between the bosonic
 pieces of the zero picture operators will contribute 
to the amplitude.  The result is
\begin{equation}
A_f = \frac{G_o^2}{\alpha'_e} \, \delta^2 \left(\sum_i k_i \right)
\pi (1-s) \frac{\cos(\pi s) + \cos[\lambda (1-s)]}{\sin(\pi s)} 
\, \xi_1 \cdot \xi_3 \tr(\zeta_2 \cdot \zeta_4). 
\end{equation}
This amplitude clearly vanishes in the limit $\theta = 2 \pi \alpha'_e$ or equivalently $\lambda = \pi$.  

Finally, we need to consider back scattering:
\[
k_1 = \left( \begin{array}{c} p \\ p \end{array} \right) \;\;\;
k_2 = \left( \begin{array}{c} e \\ -p \end{array} \right) \;\;\;
k_3 = \left( \begin{array}{c} -p \\ p \end{array} \right) \;\;\;
k_4 = \left( \begin{array}{c} -e \\ -p \end{array} \right).
\]
The result is the most complicated in the paper:
\begin{eqnarray}
\lefteqn{A_b = \frac{G_o^2}{\alpha'_e} \, \delta^2 \left(\sum_i k_i \right)
\left((1+t) B_1 \xi_1 \cdot \xi_3 \, \tr(\zeta_2 \cdot \zeta_4) +\right.} 
\nonumber \\
& & \left. (B_2 - tB_3)\xi_1 \cdot \zeta_2 \cdot \zeta_4 \cdot \xi_3 +
(B_4 - t B_5)\xi_1 \cdot \zeta_4 \cdot \zeta_2 \cdot \xi_3 \right) 
\end{eqnarray}
where
\begin{eqnarray*}
B_1 &=& B(2-u, 2-s) + 
B(-1-t, 2-s) \cos[\lambda (1-s)] +
B(-1-t, 2-u) \cos[\lambda (1-u)]\\
B_2 &=& B(2-u,-s) + 
B(1-t,-s) \cos[\lambda (1-s)] +
B(1-t,2-u) \cos[\lambda (1-u)]\\
B_3 &=& B(2-u, 1-s) -
B(-t, 1-s) \cos[\lambda (1-s)] +
B(-t,2-u) \cos[\lambda (1-u)] \\
B_4 &=& B(-u, 2-s) +
B(1-t, 2-s) \cos[\lambda (1-s)] +
B(1-t, -u) \cos[\lambda (1-u)] \\
B_5 &=& B(1-u,2-s) +
B(-t, 2-s) \cos[\lambda (1-s)] -
B(-t, 1-u) \cos[\lambda (1-u)].
\end{eqnarray*}

In the NCOS limit, the vanishing of all of these $B_i$ is equivalent to the one 
Beta function identity
\[
B(x,y) - B(z,x) \cos(\pi x) - B(z,y) \cos(\pi y) = 0
\]
where $x+y+z=1$ or equivalently here $s+t+u=2$.  This identity is the same one 
that appeared in demonstrating the vanishing of the back scattering of a bosonic 
tachyon off of a massless scalar.

\section{Decay into Wound Closed Strings}

In the uncompactified NCOS theory, the reason why the closed strings
decouple from the open strings is largely kinematic:  
The open strings simply do
not have enough energy to produce closed strings in the NCOS scaling
limit, $\alpha' = \alpha'_e(1-E^2) \rightarrow 0$ and $\alpha'_e$ held
fixed.
%
However, when we compactify along the 1 direction with radius $R$, the
wound closed strings with winding number $w$ and $n$ quanta of momentum in
the compactified direction satisfy the onshell condition (see for example
\cite{Polchinski}):
\[
-\alpha' (p_0)^2 - 2p_0 E w R + \frac{(wR)^2}{\alpha'_e} + \alpha' (n/R)^2
+ 2(N_L + N_R) + \alpha'_e p_\perp^2 = 0,
\]
where $p_\perp$ is the transverse momentum, along with a level matching
condition $N_L - N_R = wn$.  In the NCOS scaling limit, strings with
nonzero $w$ do not acquire infinite energy, and, as shown in
\cite{Klebanov}, the dispersion relation becomes:
\[
p_0 = \frac{wR}{2 \alpha'_e} + 
\frac{\alpha'_e}{2wR} p_\perp^2 + \frac{N_L + N_R}{wR}.
\]
If we consider the case of the wound graviton, $N_L = N_R = 0$ and $n=0$,
then, provided we make $R$ small enough, $wR < 2\sqrt{l\alpha'_e}$, energy
conservation will allow the decay of any massive open string state at
level $l$ into a wound graviton.  Roughly speaking, this condition is
equivalent to saying that a massive open string which is long enough to
wrap around the compactified direction $w$ times 
can turn, by joining its ends,
into a wound closed string with winding number $w$.

In pure open
string amplitudes, we were able to carry out the NCOS scaling limit at the
level of Green's functions and vertex operators.  Once we include wound
states, we have to first work away from the scaling limit
and take it only at the end.  
On the disk, we may use the doubling trick to express the $X(z,\bar{z})$
field in terms of its holomorphic part only~\cite{Hashimoto}.  In
particular,
\[
X^\mu (z, \bar{z}) = X^\mu(z) + {(DM^{-1})^\mu}_\nu X^\nu(\bar{z}). 
\]
where the
$M$ matrix comes from the boundary conditions imposed by the electric
field on the open strings.  In the directions transverse to the D-string,
$M$ is the identity, while parallel to the D-string
\[
{M^\alpha}_\beta  = \frac{1}{1-E^2} \left( \begin{array}{cc}
1+E^2 & -2E \\
 -2E & 1+E^2  \end{array} \right). 
\]
The $D$ matrix is the identity in the
directions parallel to the D-string and minus the identity in the
directions transverse to the D-string.  A lengthier discussion of these
$M$ and $D$ matrices can be found in \cite{Hyun}.   

With this
doubling trick in mind, the Green's functions we will use are the usual
\begin{eqnarray*}
\langle X^\alpha(z) X^\beta(z') \rangle &=& 
-\frac{\alpha'}{2} \eta^{\alpha\beta} \ln(z-z') \\
\langle X^i(z) X^j(z') \rangle &=&
-\frac{\alpha'_e}{2} \delta^{ij} \ln(z-z') \\
\langle \psi^\mu (z) \psi^\nu (z') \rangle &=& \frac{\eta^{\mu\nu}}{z-z'}
\ ,
\end{eqnarray*}
where $\alpha, \beta =0,1$ and $i,j=2, \ldots, 9$.
As we are considering a two point function, the phases will not 
contribute.  

Let us consider an open string state on the leading Regge trajectory
decaying from rest. 
It has no momentum in the spatial direction: $k_1=0$. Although  
we have done the calculation for more general polarization tensors, for
clarity we will present the results only for states polarized transversely to
the D-string, i.e. the states on the leading SO(8)
Regge trajectory. In section 2 we showed that all such  
states are stable in the uncompactified theory. Now we demonstrate that,
when the spatial direction is compact, sufficiently heavy such states
can decay into wound closed strings. It is enough to consider winding
states of the graviton polarized transversely to the D-string.
We take the open string vertex operator in the zero picture 
while the wound graviton will be in the $(-1,-1)$ picture
to saturate the ghost number on the disk: 
\footnote{This normalization
of the winding vertex operator produces finite amplitudes on a sphere,
with effective interaction strength of the winding states $\sim G_o^2$.
We are grateful to M. Krasnitz for helpful discussions of this issue.}
\begin{eqnarray}
V_{m,0} &=& G_o \left(\frac{2}{\alpha'_e}\right)^{\frac{l+1}{2}} 
\frac{1}{\sqrt{l!}}
\xi_{i j_1 \ldots j_l} 
\left( \partial X^i \partial X^{j_1} \ldots \partial X^{j_l} + \right. 
\nonumber \\
&& 
- \alpha'_e l \; \psi^i \partial \psi^{j_1}  
\partial X^{j_2} \ldots \partial X^{j_l} +
\nonumber \\
&&
\left.
+ i \alpha'_e (k \cdot \psi ) \psi^i 
\partial X^{j_1} \ldots \partial X^{j_l} \right) 
e^{i k \cdot (1+M^{-1}) \cdot X} (y) 
\label{Rtraj} \\
V_{g,-1,-1} &=& G_o^2
\epsilon_{i j} \left( e^{-\phi} \psi^j  
e^{ip_R \cdot DM^{-1} \cdot X} \right) (\bar{z})
\left( e^{-\phi} \psi^i e^{ip_L \cdot X} \right) (z)
\ .\end{eqnarray}
%
The polarization tensors $\xi_{i j_1 \ldots j_l}$ and
$\epsilon_{i j}$ are completely symmetric and traceless.
Because we have compactified, we need to use the right and left momenta
$p_{0L} = p_{0R} = p_0 + EwR/\alpha'$, $p_{1L,R} = \pm wR / \alpha'$ and
$p_{iL} = p_{iR} = p_i$.  On shell, $-k^2 = (k_0)^2 = l/\alpha'_e$, and
energy conservation means that $k_0 = -p_0$. 

In calculating the amplitude, only the second term in (\ref{Rtraj}) will
contribute to the contractions. The amplitude is 
coordinate independent, as we would expect from $SL_2(R)$ invariance:
\begin{eqnarray}
\sqrt{l!} \; {\mathcal A} (k,\xi;p,\epsilon) 
&\sim &
2 l\frac{G_o}{\alpha'_e} \left(\frac{\alpha'_e}{2}\right)^{(l-1)/2} 
%
\xi_{i j_1 \ldots j_l} \epsilon^{i j_1} p^{j_2} \ldots p^{j_l}
\end{eqnarray}
  In the NCOS limit, the amplitude is
finite.  Therefore, it is indeed possible for sufficiently heavy
leading Regge trajectory states to decay into wound gravitons. 
 We conjecture that
the same holds true more generally:  Even if decay into lighter open
string states is prohibited, any massive open string state can decay into
lighter wound closed string states.

\section{Remarks}

The general argument presented in \cite{Klebanov} 
along with the many open string scattering amplitudes 
calculated both in \cite{Klebanov} 
and in this paper show 
that the massless open string modes do indeed decouple from the rest 
of the open string spectrum.  As we saw here, one interesting consequence 
of the decoupling is that a number of the massive open string modes 
become stable in the uncompactified theory.  
It is interesting to speculate as to what these stable 
states correspond in the dual U(N) gauge theory.  One possibility is a 
set of nontopological solitons: localized lumps of finite energy.
Roughly speaking, such a lump is a section of an individual D-string
which becomes liberated from the bound state and spins around it
with sufficiently high SO(8) angular momentum. In the compactified theory,
sufficiently highly excited such states may decay into Higgs branch
excitations; there is enough energy for the entire D-string to become
liberated. This process is dual to the transition from an open to
a wound closed string in NCOS that was studied in section 4.

For large $N$ the dual NCOS is weakly coupled.  Hence the
stable states are predicted to have $m^2 = n g_{YM}^2/N^2$, where
$n$ are integers.
For finite $N$ further corrections in $G_o^2=1/N$ should appear, and the
spectrum will no longer be linear. It seems likely, however, that
some of the states will remain stable for all $N>1$.
Perhaps these stable states may be found through numerical simulations of
the gauge theory with an electric flux line. 

\section*{Acknowledgments}
We are grateful to M. Krasnitz for many useful discussions.  
The work of C.H. was supported in part by a NDSEG Fellowship.  
The work of I.R.K. was supported in part by the NSF grant PHY-9802484 and
by the James S. McDonnell Foundation Grant No. 91-48.


\begin{thebibliography}{99}

\bibitem{Klebanov} I.~Klebanov and J.~Maldacena, ``$1+1$ Dimensional NCOS and 
its U(N) Gauge Theory Dual,'' hep-th/0006085.

\bibitem{Seiberg} N.~Seiberg, L.~Susskind, and N.~Toumbas, ``Strings in 
background electric field, space/time non-commutativity and a new non-critical 
string theory,'' hep-th/0005040.

\bibitem{Gopak} R.~Gopakumar, J.~Maldacena, S.~Minwalla, and 
A.~Strominger, ``S-duality and noncommutative gauge theory,'' 
hep-th/0005048.

\bibitem{Witten} E.~Witten, ``Bound States of Strings and p-Branes,''
{\it Nucl. Phys.}, {\bf B460}  (1996) p 335, hep-th/9510135.

\bibitem{Gukov} S.~Gukov, I.~Klebanov, and A.~Polyakov, ``Dynamics of (n,1) 
strings,'' {\it Phys. Lett.} {\bf B423} (1998) p 64, hep-th/9711112. 

\bibitem{Verlinde} H.~Verlinde, ``Matrix String Interpretation of the Large N 
Loop Equation,'' hep-th/9705029.

\bibitem{Gopak2} R.~Gopakumar, S.~Minwalla, N.~Seiberg, 
and A.~Strominger, ``(OM) Theory in Diverse Dimensions,'' hep-th/0006062.

\bibitem{Bala} V.~Balasubramanian and I.~Klebanov, ``Some Aspects of Massive 
World-Brane Dynamics,'' {\it Mod. Phys. Let.} {\bf A11} (1996) p 2271, 
hep-th/9605174.

\bibitem{FultonHarris} W.~Fulton and J.~Harris, {\it Representation Theory: A 
First Course}, Springer, 1991. 

\bibitem{RCKing} R.~C.~King, ``Modification rules and products of irreducible 
tensor representations of the unitary, orthogonal and symplectic groups,'' {\it 
J. Math. Phys.} {\bf 12} (1971).

\bibitem{FT} E.~Fradkin and A.~Tseytlin, ``Nonlinear 
Electrodynamics From Quantized Strings,'' {\it Phys. Lett.} {\bf 163B} (1985) p 
123; C.~Callan, C.~Lovelace, C.~Nappi, and S.~Yost, 
``String Loop Corrections to Beta Functions,'' {\it Nucl. Phys.} 
{\bf B288} (1987) p 525; A.~Abouelsaood, 
C.~Callan, C.~Nappi, an S.~Yost, ``Open Strings In Background Gauge Fields,'' 
{\it Nucl. Phys.} {\bf B280} (1987) p 599.

\bibitem{SeiWit} N.~Seiberg and E.~Witten, ``String Theory and Noncommutative 
Geometry,'' hep-th/9908142. 

\bibitem{Polchinski} J.~Polchinski, {\it String Theory}, vol. 1, Cambridge Univ. Press, sec. 8.4.

\bibitem{Hashimoto} A.~Hashimoto and I.~Klebanov, ``Scattering of Strings from D-branes,'' hep-th/9611214.

\bibitem{Hyun} S.~Hyun, Y.~Kiem, S.~Lee, and C.-Y.~Lee, 
``Closed Strings Interacting with Noncommutative D-branes,'' 
{\it Nucl. Phys.} {\bf B569} (2000) p 262, hep-th/9909059.

\end{thebibliography}
\end{document}